%% file: 4D-DSS_network.tex
\documentclass[journal]{IEEEtran}
\usepackage{ifpdf}

\ifCLASSINFOpdf
   \usepackage[pdftex]{graphicx}
\else
   \usepackage[dvips]{graphicx}
\fi
\usepackage{mathtools}
\usepackage{amsmath}
\usepackage{algorithmic}

\usepackage{array}

\usepackage{booktabs}
\usepackage{graphicx}
\usepackage{subfigure}
\usepackage{float}

\ifCLASSOPTIONcompsoc
  \usepackage{subfig}
\else
  \usepackage{subfig}
\fi
\begin{document}
%
\title{Automatic Left Ventricular Cavity Segmentation via Deep Spatial Sequential Network in 4D Computed Tomography Studies}
%
%
%
\author{Yuyu Guo,
        Lei Bi,
        Zhengbin Zhu, 
        David Dagan Feng,~\IEEEmembership{Fellow,~IEEE,}
        Ruiyan Zhang,
        Qian Wang,~\IEEEmembership{Member,~IEEE,}
        and~Jinman Kim,~\IEEEmembership{Member,~IEEE,}
\thanks{Y. Guo and Q. Wang are with the School of Biomedical Engineering, Shanghai Jiao Tong University, 200240 China,
e-mail: wang.qian@sjtu.edu.cn.}
\thanks{L. Bi, D. Feng and J. Kim are with the School of Computer Science, University of Sydney, NSW 2006 Austrilia, e-mail: jinman.kim@sydney.edu.au}
\thanks{Z. Zhu and R. Zhang are with Ruijin Hospital, Shanghai Jiaotong University School of Medicine, 200025 China}}
\markboth{}%
{}
%



\maketitle

\begin{abstract}
Automated segmentation of left ventricular cavity (LVC) in temporal cardiac image sequences (multiple time points) is a fundamental requirement for quantitative analysis of its structural and functional changes. Deep learning based methods for the segmentation of LVC are the state of the art; however, these methods are generally formulated to work on single time points, and fails to exploit the complementary information from the temporal image sequences that can aid in segmentation accuracy and consistency among the images across the time points. Furthermore, these segmentation methods perform poorly in segmenting the end-systole (ES) phase images, where the left ventricle deforms to the smallest irregular shape, and the boundary between the blood chamber and myocardium becomes inconspicuous. To overcome these limitations, we propose a new method to automatically segment temporal cardiac images where we introduce a spatial sequential (SS) network to learn the deformation and motion characteristics of the LVC in an unsupervised manner; these characteristics were then integrated with sequential context information derived from bi-directional learning (BL) where both chronological and reverse-chronological directions of the image sequence were used. Our experimental results on a cardiac computed tomography (CT) dataset demonstrated that our spatial-sequential network with bi-directional learning (SS-BL) method outperformed existing methods for LVC segmentation. Our method was also applied to MRI cardiac dataset and the results demonstrated the generalizability of our method. 
\end{abstract}


%
\IEEEpeerreviewmaketitle

\section{Introduction}
%
%
%
%
\IEEEPARstart{H}{eart} failure is a global problem \cite{savarese2017global,ponikowski2014heart}. A study in 2014 by Ponikowski et al. \cite{ponikowski2014heart}, projected that approximately 26 million people are living with heart failure (HF). Accordingly, HF expenditure is considerable and expected to increase dramatically with the ageing population \cite{van2016epidemiology}. Early detection is thus an effective strategy to reduce the mortality and morbidity associated with HF \cite{minicucci2011heart}. In particular, measurement of the shape changing in the left ventricular (LV) structure is important as it can be used in the early detection and analysis of complication of myocardial infarction \cite{konstam2011left}.

\begin{figure}[]

\centering
\includegraphics[width=0.5 \textwidth]{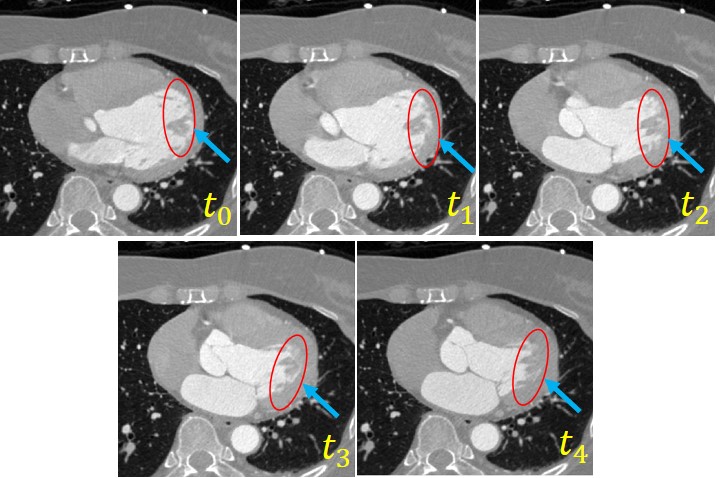}

\caption{Comparison of LVC shape changes during cardiac contraction period using 4D-CT. The images represent the LVC from end-diastole (ED – t0) and end-systole (ES – t4) which shows the deformation caused in LVC contraction from blood pumping out during the systole phase. The red circles and blue arrows are used to indicate the inference of papillary muscles and the trabeculae, which is responsible for the boundary of the LVC in ES period becoming blurry.}
\label{fig:1}
\end{figure}

For LV assessment, multiple cardiac imaging modalities, including computed tomography (CT), X-rays, ultrasound (echocardiogram) and magnetic resonance (MR) imaging, are routinely used for diagnosis and ongoing assessments \cite{celebi2010current,chahal2010clinical}. In particular, four-dimensional (4D) 4D-CT, which sequentially captures the cardiac cycle from diastole to systole phase (as exemplified in Fig. \ref{fig:1}), allows for non-invasive imaging of cardiac structures with high contrast and high resolution \cite{tavakoli2014cardiac}. With 4D-CT, both the structural and functional properties of LV motion are captured and enables to quantitatively model the cardiac cycle for abnormalities \cite{bardinet1996tracking,mihalef2010patient}. 

To model the LV, left ventricle cavity (LVC) structure is required to be segmented at individual time-points of the cardiac cycle, and the segmentation results can then be used for the derivation of the clinical indices such as the ejection fraction, wall thickness and motion \cite{tavakoli2014cardiac,mihalef2010patient,gao20114d}, which are the commonly used attributes in the early diagnosis of myocardial infarction. However, current approaches to LV modelling in the clinical workflow is reliant on tedious, time-consuming and non-reproducible manual segmentation process \cite{de2011comparison,juergens2006left,bernard2018deep}, where volume sequences need to be segmented incorporating spatial and temporal consistency.

Existing automated segmentation methods for LVC, unfortunately, are generally optimized for a single time point of a cardiac cycle \cite{wang2014direct,oktay2017anatomically,zheng20183}. These methods, therefore, do not leverage the cardiac imaging sequences which comprises of both spatial and temporal information that can be used in combination towards more consistent, robust and, accurate segmentation results. For instance, these methods are subject to performance degradation in difficult cardiac phases such as in ES where the left ventricle deforms to the smallest irregular shape and the boundary between the blood chamber and myocardium becomes inconspicuous due to muscle interference such as the papillary muscles and the trabeculae (see Fig. \ref{fig:1}). Among the few works that have been specifically designed for temporal LVC segmentation \cite{xue2017full,yan2018left}, they were for 2D image sequences and for MR images (cardiac cine or tagged MR imaging). The segmentation algorithms optimized for such MR cardiac studies, although demonstrated the values from the use of temporal information, the imaging characteristics are distinctly different in imaging characteristics to that of 4D-CT, including lower spatial resolution (poorer slice-thickness), temporal resolution (number of time-points) and, reduced contrast / texture differentiability of the LVC.

\subsection{Related works}
There is paucity of research literature on automated LVC segmentation from 4D-CT images. We therefore included works on the segmentation of other cardiac structures such as the left ventricular myocardium (LVM) and right ventricular cavity (RVC) in both CT and MR images. We partitioned the related works into two categories: (1) conventional segmentation methods and (2) deep learning based segmentation methods. 

Conventional methods make use of local image features in the form of shape models \cite{mahapatra2013cardiac,heimann2009statistical}, textures \cite{eslami2013segmentation,huang2007dynamic}, edges \cite{sethian1999level,chenoune2005segmentation} and regions features \cite{zheng2008four}, with graph models \cite{lee2010automatic,mitchell20023} or classifiers \cite{lorenzo2004segmentation}. The shape model methods \cite{paragios2003level,lynch2008segmentation,zhuang2010registration} work by building a model (via atlas registration or deformation) with priori-knowledge of the cardiac structures. The built model was then applied to segment the cardiac structures based on image features such as intensity or texture. Unfortunately, these methods are highly reliant on accurate registration and prior knowledge, e.g., shape priors, and therefore these methods do not perform well in cases where there exists large deviation of the shape from the model. Other common approaches include graph models and classifier-based methods. These methods were designed to extract features such as textures \cite{kotu2011segmentation} and edge-based features \cite{somkantha2010boundary}, for use in a graph models and/or classifiers such as random walkers \cite{paragios2003level} and support vector machines \cite{bai2015multi,karim2016evaluation} to detect and segment the cardiac structures. In these methods, the reliance on using local features meant that they constituted limited capability in capturing image-wide semantic information. In addition, performances for these methods depended on correctly tuning a large number of parameters and effective pre-processing techniques such as image de-noising and model constructions, which may limit their generalizability to a different dataset.

Recently, deep learning based methods have achieved the state-of-the-art performances in various segmentation tasks on medical images \cite{cciccek20163d, long2015fully, lin2017refinenet}. Such success is primarily attributed to these methods having the capability to leverage large training datasets to derive feature representations that carry high-level semantics as well as low-level appearance information in a hierarchical manner. Motivated by this success, many investigators have translated deep learning methods for cardiac structures segmentation \cite{avendi2016combined,tan2017convolutional,zheng20183,vigneault2018omega,tao2018deep}. Most of these methods are optimized for 2D images \cite{tan2017convolutional,wolterink2017automatic}, 2.5D (multiple views) images \cite{mortazi2017multi,luo2017multi} and 3D images \cite{zheng20183}, all from a single time point for LVC or RVC segmentation using MR modality \cite{baumgartner2017exploration}. These single time-point methods made use of deep neural networks with various cardiac shape constraints (or prior knowledge) e.g., deformable models, or ensemble of 2D and 3D segmentation results \cite{isensee2017automatic,jang2017automatic}. For instance, Isensee et al. \cite{isensee2017automatic} presented an ensemble of U-Net architectures for segmentation of the cardiac structures with cine-MR and achieved the best result on Automated Cardiac Diagnosis Challenge (ACDC) dataset \cite{bernard2018deep}. Deep learning methods optimized for 4D data has demonstrated improved performance compared to a single-point counterpart. Xue et al. \cite{xue2017full} integrated a recurrent neural network (RNN) module into CNN for LV segmentation with MR; however it relies on training massive learnable RNN parameters which are difficult to converge and require a mass of training data. Yan et al. \cite{yan2018left} proposed to use optical flow to model the cardiac motion information and was used to aid in LVC segmentation on MRI images. In this approach, the optical flow information was derived by using paired image slices across adjacent time-points. However, this approach was simply paired image slices using index numbers which is not realistic and was only demonstrated for 2D images.


\subsection{Our Contributions}
In this study, we present an automated LVC segmentation method for 4D-CT images. We introduce a spatial-sequential network with bi-directional learning (SS-BL) technique, where we propose to leverage the complementary data in the cardiac cycle image sequences for segmentation. Compared to the existing methods, we have the following contributions:

\begin{itemize}
\item[$\bullet$] A 3D spatial-sequential (SS) network is proposed to learn the deformation and motion information of the cardiac structures in 4D images, in an end-to-end manner. SS network is based on the prediction of cardiac motion in a 3D image space during a complete cardiac cycle and is used to guide the LVC segmentation. A deformation consistency loss is further introduced to constrain the learning of the latent feature space, and therefore enforcing the motion reversibility among the image sequences to result in consistent transition among the sequential segmentation results.

\begin{figure*}[h]
\centering
\includegraphics[width=0.8 \textwidth]{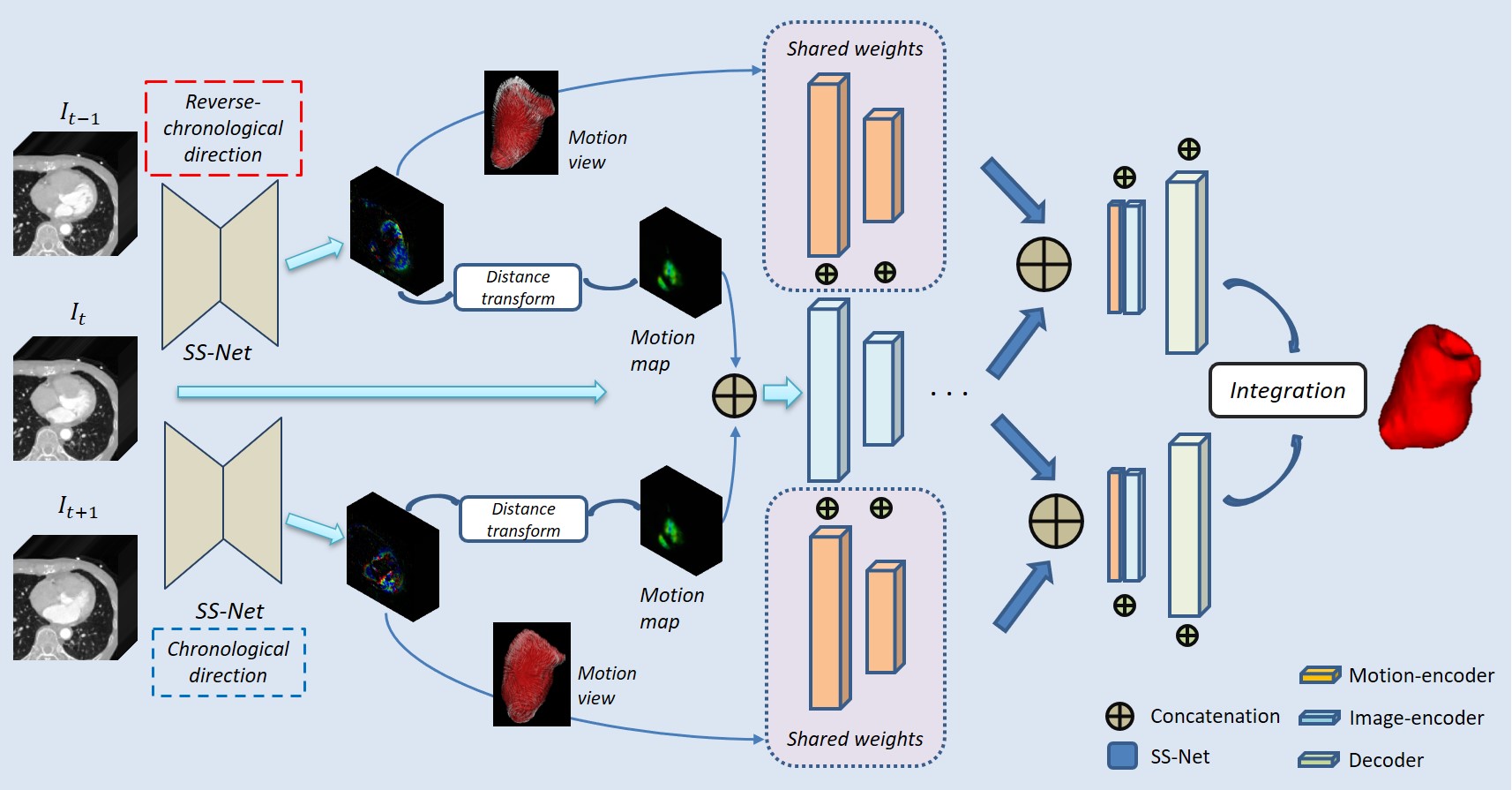}
\caption{ Overview of our spatial-sequential network with bi-directional learning (SS-BL) method for temporal cardiac image segmentation.}
\label{fig:2}
\end{figure*}

\item[$\bullet$] We propose a bi-directional learning (BL) approach to learn the sequential-context information to refine the segmentation results of the cardiac structure boundaries. We innovate in the utility of complementary segmentation results produced in bi-directional time sequences (include two single-directional learning (SL) - chronological and reverse-chronological directions) to enhance the agreements between the segmentation results of the adjacent time-points. This technique ensures that the appearance of the cardiac structures is spatially consistent in both chronological and reverse-chronological directions, and thus resulting in improved segmentation results of the boundaries.

\item[$\bullet$] To our knowledge, we present the first deep learning based segmentation of the LVC that fully leverage both the spatial and temporal information available in 4D-CT images. For broader applications to cardiac research, we demonstrate the generalizability of our SS-BL to segment different cardiac structures, including LVM and RVC in 4D MR cardiac images.
\end{itemize}

The rest of the paper is organized as follows. Section 2 describes our methods and materials. Section 3 presents the experimental results on comparing our method to the existing state-of-the-art. This is followed by discussions in Section 4, and conclusions are made in Section 5.

\section{Methods and Materials}

\subsection{Overview of the Framework}

The overview of our SS-BL method is illustrated in Fig. \ref{fig:2}. Our method consists of two major components: (i) a 3D spatial-sequential network (SS) for unsupervised motion estimation based on CT images in adjacent time points; and (ii) a bi-directional learning (BL) approach to integrate the segmentation results in both chronological and reverse-chronological direction.

\subsection{Spatial-Sequential Network}

\begin{figure*}[h]
\centering
\includegraphics[width=0.8 \textwidth]{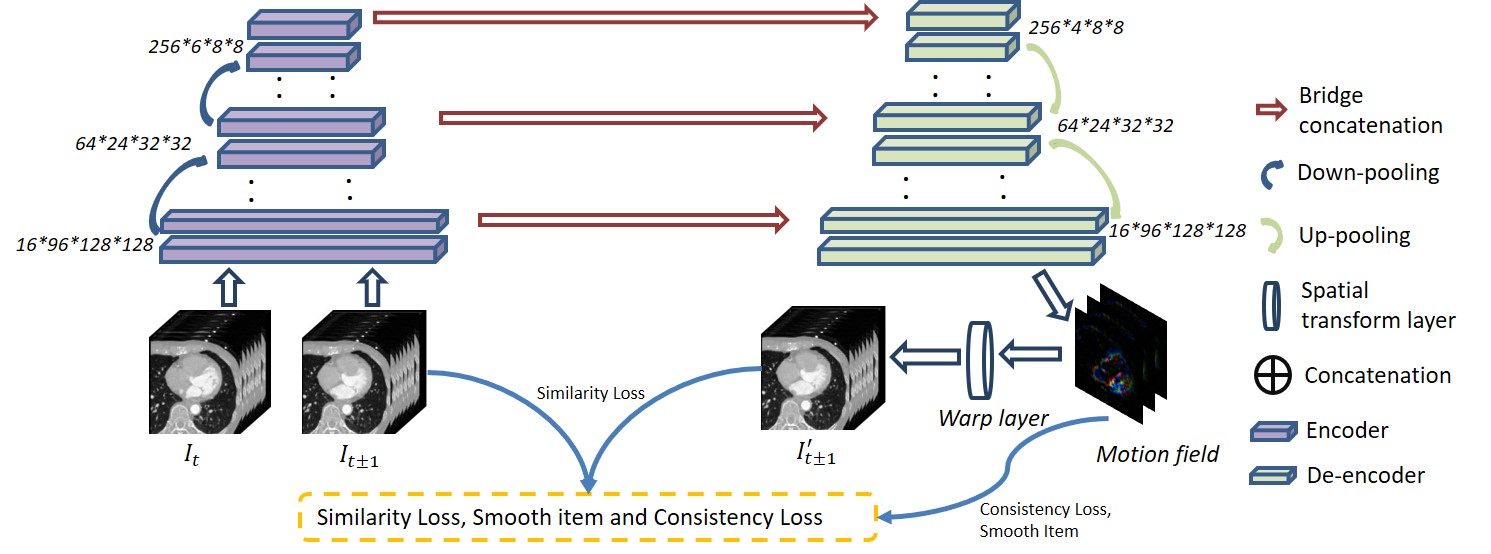}
\caption{Overview of our spatial-sequential network.}
\label{fig:3}
\end{figure*}

Our SS network was used to learn the motion field of cardiac during the whole cardiac cycle and was inspired by optical flow based methods \cite{meister2018unflow,balakrishnan2018unsupervised}. In contrast, we attempted to model the cardiac motion information in a volumetric space and in an unsupervised manner, as illustrated in Fig. \ref{fig:3}. We used a modified 3D U-net architecture with two input variants (adjacent image volumes) for learning and deriving the deformations. The convolutional layers extracted the features hierarchically to compare the structural difference between the two input volumes and then to estimate the intensity correspondence related to the spatial information.

The U-net architecture was used in SS-Net is efficient to expand the receptive field which is able to cover the cardiac motion. All the convolutional kernels were set to \textit{$3\times3\times3$} and through 4 times downsampling to enlarge the receptive field. The skip connections were also applied to fuse low level structural features with high level semantic features. The output of SS-Net \textit{$\varphi$} indicates the spatio-temporal velocity field between two inputs cardiac volumes.

We used a spatial transformation layer \textit{$T$} to warp the two input variants of SS-Net with a bi-linear interpolation. It enabled to achieve the voxel coordinate transformation by obtaining the motion field. The aim of the SS-Net is to minimize the difference between the transformed and the image \textit{$I_{t}$} at time point \textit{$t$}. To achieve this, we defined a content-based loss function as:

\begin{equation}\label{eq:1}
    \ell{oss}_{motion}=argmin(\sum^{n}_{i=1}\left| T(I_{t\pm{1}}| \varphi_{t\pm{1}})-I_{t}\right|+\psi+\sigma))
\end{equation}

Where \textit{$I_{t}$} and \textit{$I_{t-1}$} are two adjacent image volumes at time point \textit{$t$} and \textit{$(t-1)$}. \textit{$|\cdot|$} is a \textit{$L_{1}$} regulation. \textit{$\varphi$} is the output 3D motion field and \textit{$T(\cdot)$} is a spatial transformation layer.  \textit{$n$} represents all the voxels within the image volume. To further improve the content of the output deformation field, we added smoothness term \textit{$\psi$} and consistency term \textit{$\sigma$}. The smoothness term was defined as:

\begin{equation}\label{eq:2}
    \psi=\dfrac{1}{n-1}(\dfrac{\partial{\varphi(x,y,z)}}{\partial{x}}+\dfrac{\partial{\varphi(x,y,z)}}{\partial{y}}+\dfrac{\partial{\varphi(x,y,z)}}{\partial{z}})
\end{equation}

Where \textit{$\varphi(x,y,z)$} represents the voxel values at a spatial location of the deformation field. As the deformation between two input image is reversible, we introduced a consistency term \textit{$\sigma$} which is defined to minimize the difference between the deformation field in both chronological and reverse-chronological of cardiac motion. For instance, the deformation from \textit{$I_{A}$} to \textit{$I_{B}$} is \textit{$\varphi_{AB}$}, and the deformation from \textit{$I_{B}$} to \textit{$I_{A}$} is \textit{$\varphi_{BA}$}, the following equation can be established:

\begin{equation}\label{eq:3}
    I_{B}=T(I_{A}|\varphi_{AB})\quad\text{or}\quad I_{A}=T(I_{B}|\varphi_{BA})
\end{equation}

Base on prior that deformation is reversible which means we can also get \textit{$I_{A}$} or \textit{$I_{B}$} from \textit{$\varphi_{AB}$} or \textit{$\varphi_{BA}$}, then we can have:

\begin{equation}\label{eq:4}
    \varphi_{BA}=-T(\varphi_{AB}|\varphi_{AB})\quad\text{or}\quad \varphi_{AB}=-T(\varphi_{BA}|\varphi_{BA})
\end{equation}

Thus, the consistency term \textit{$\sigma$} can be defined as: 

\begin{equation}\label{eq:5}
    \sigma=\sum_{i=1}^{n}|T(\varphi|\varphi)+\varphi^{-1}|
\end{equation}

where \textit{$\varphi^{-1}$} represents the deformation between \textit{$(t-1)$} and \textit{$t$}, which is the reverse deformation field of \textit{$\varphi$}. 

\subsection{Bi-directional Learning for Cardiovascular Image Segmentation}

As shown in Fig. \ref{fig:2}, we utilize a 3D M-Net architecture \cite{jang2017automatic}, where both the motion information and the intensity image were used in two independent branches to guide the cardiac segmentation. We model a segmenting function \textit{$G_{\theta}(I,\varphi_{\pm{1}}=L$} using our SS-SL network, where \textit{$L$} indicates the segmentation result and \textit{$\theta$} are the learnable parameters of \textit{$G$}. The SS-BL is composed of three types of modules – SS-Net, two individual encoder branch and decoder (SS-SL). For the encoder module, two input branches (image-based branch and motion-based branch) were designed to extract deformation features and image  structural features separately. The image-based branch including the intensity image, moreover, we also add a motion-based distance map generated by the deformation field from SS-Net to guide the image-based encoder to pay attention to the cardiac region. For the motion-based branch, it includes the deformation field \textit{$\varphi$} from target image with its adjacent image (see \ref{fig:3}). The spatial movement information suppresses distracted structure but captures more boundary structural information and object location within a certain vision window, while the intensity image provides more details of the cardiac shape structure. As the heart beats is a four-dimensional movement, which accompanied by twisting action during contraction to relaxation, we applied spatial transform prediction module on 3D space to fit the real motion of cardiac. In practice, predicting the deformation field of the paired images to obtain the motion information firstly, then input the 3D volume image to be segmented and deformation field to the segmentation network.

\begin{figure}[h]
\centering
\includegraphics[width=0.45 \textwidth]{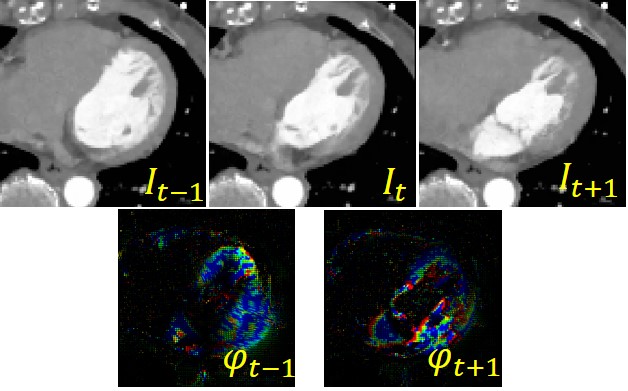}
\caption{The bi-directional spatiotemporal motion field.}
\label{fig:4}
\end{figure}

The cardiac motion has a sequence on a time axis which has two move directions in each time-point as Fig. \ref{fig:4}. To keep the continuous and smoothness of motion in timeline, we proposed a bi-directional learning to include both chronological and reverse-chronological in spatial sequential motion to guide cardiac segmentation. For SS-BL network, modelling the deformation between \textit{$I_{t}$} and its chronological time-phase image \textit{$I_{t-1}$}, then combined motion information and intensity information to predict segmentation result \textit{$I_{t}^{f}$} based on chronological motion. Similarly, generate another segmentation result \textit{$I_{t}^{r}$} based on \textit{$I_{t}$} and its reverse-chronological direction time-phase image \textit{$I_{t+1}$}. Fusing these two segmentation results \textit{$I_{t}^{f}$} and \textit{$I_{t}^{r}$} to obtain the combined segmentation result \textit{$I_{t}$}.

\begin{equation}\label{eq:6}
    L_{t}=G(I_{t},\varphi_{t-1}|\theta)+G(I_{t},\varphi_{t+1}|,\theta)
\end{equation}

\subsection{Materials}
Our 4D cardiac CT (4D-C-CT) imaging dataset was acquired at the department of Cardiology, Ruijin hospital affiliated to the school of medicine, Shanghai Jiao Tong University, Shanghai, China. The dataset consists of 18 patient data, each temporal study containing CT scans acquired at 10 consecutive time-points (t0 - t9) across a single cardiac cycle. Our cycle image concludes two phases, systole and diastole, which begin at systole (t0-t4) and then diastole occurs (t5-t9), respectively. The imaging has a high-resolution ranging from 0.32 to 0.45 mm in intra-slice (x- and y-resolutions) and from 0.37mm to 0.82mm in inter-slice (z-resolution). The ground truth of left ventricular cavity (LVC) was manually segmented for all 10 time-points by an experienced technician and was then confirmed by a senior clinician in a slice-by-slice manner. The dataset was processed by contrast normalization to highlight the soft-tissues. To reduce the pixel variations among the patient studies, \textit{$0.5\%$} of the pixels with the highest intensity values were excluded \cite{jang2017automatic}. Patch size was set to \textit{$128\times128\times96$}.

\subsection{Implementation Details}
We implemented our SS-BL method with Pytorch library and was trained on a 11GB Nvidia 1080Ti GPU. SS Network was trained for 200 epochs and SS-BL for 100 epochs. Both networks were trained with an Adam optimizer at a learning rate of 0.0001 and a batch size of 1. It took on average of 20 hours to train the SS network and an additional average of 8 hours to train the SS-BL.

\section{Experiment and Results}
\subsection{Experiment Setup}
We conducted the following experiments: (i) comparison of our SS-BL with the state-of-the-art segmentation methods; (ii) compartmental analysis of our SS-BL and, (iii) analysis of the variation in the number of time-point phase sequences (temporal information). For the 4D-C-CT dataset, 6-fold cross-validation was applied, with each fold consisting of randomly selected 15 training and 3 testing data. Comparison methods included: (1) 3D M-Net – a similar network architecture to \cite{jang2017automatic} but modified to work in 3D. This network was considered as the baseline; (2) 2D / 3D U-Net \cite{ronneberger2015u} – classical medical image segmentation network; (3) OF-Net \cite{yan2018left} – optical flow guided LV segmentation network optimized for cine MR and, (4) Ensemble-CNN (E-CNN) \cite{isensee2017automatic} – 2D / and 3D CNN combined for LVC, LVM and RVC segmentations. This is the state-of-the-art cardiac segmentation method for MR cardiac images. Although both OF-Net and E-CNN were designed for MR, because they were designed for cardiac image segmentation, we suggest that they represent fair comparison methods even when applied to CT.

\subsection{Evaluation Metrics}
The standard cardiovascular image segmentation evaluation metrices were used for comparison including: Dice similarity (Dsc.), Jaccard Index (Jac.) and Hausdorff distance (HD). They are defined as:

\begin{equation}
\label{eq:4}
    Dsc(A, B)=\dfrac{2\cdot(A\cap{B})}{A + B} 
\end{equation}

\begin{equation}\label{eq:5}
    Jac(A, B)=\dfrac{A\cap{B}}{A\cup{B}}
\end{equation}

\begin{equation}\label{eq:6}
    Hd(A, B)= \max\bigg{(}\max\limits_{p\in A}\Big{(}\min\limits_{q\in B} d(p,q)\Big{)}, \\
      \max\limits_{q\in B}\Big{(}\min\limits_{p\in A} d(p,q)\Big{)}\bigg{)}
\end{equation}

where Dsc. and Jac. measure the overlap ratio between the segmentation results (\textit{$A$}) and the ground truth annotation (\textit{$B$}). \textit{$HD$} measures the distance between the two volumes with \textit{$d$} denoting the Euclidean distance, while \textit{$p$} and \textit{$q$} are the two volumes.

\subsection{Comparison with the state-of-the-art methods}%

\begin{figure*}[h]
\centering
\includegraphics[width=0.9 \textwidth]{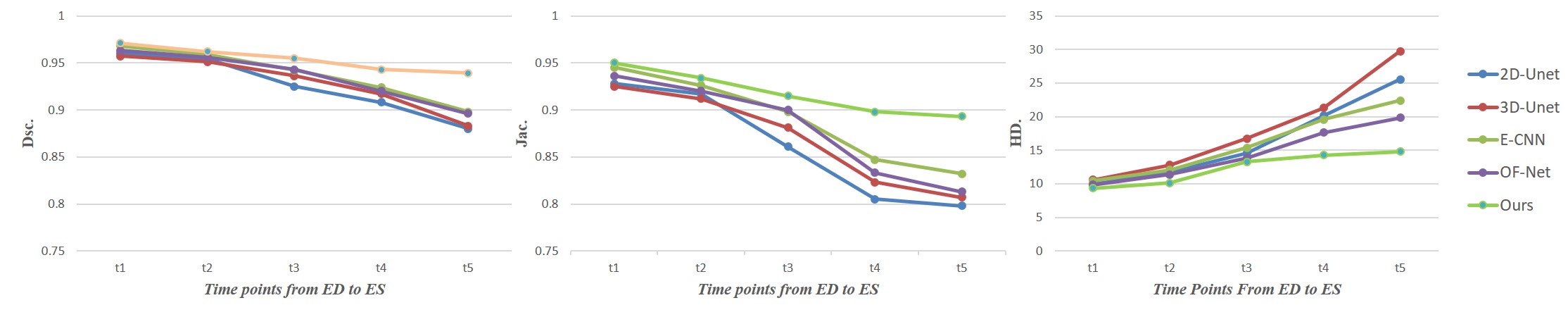}
\caption{LVC segmentation results among the individual time-points for all the comparison methods.}
\label{fig:5}
\end{figure*}

Table \ref{table:1} presents the overall results from all the segmentation methods. Fig. \ref{fig:5} plots the segmentation results at individual time-points, from the systole to diastole phases (5 time-points are shown which were selected as every second time-points from a total of 10 phases in a cardiac cycle from ED to ES). These results demonstrate that our SS-BL outperforms all other methods. Fig. \ref{fig:7} represent the qualitative segmentation results from all the comparison methods.

\begin{table}[h]
\centering
\caption{LVC Segmentation results on the 4D-C-CT dataset, where Bold represents the best results.}
\begin{tabular}{*7c}
\toprule
{} &  \multicolumn{2}{c}{Dsc.} & \multicolumn{2}{c}{Jac.} & \multicolumn{2}{c}{HD.}\\
\midrule
{}   & Mean   & Std    & Mean   & Std   & Mean   & Std\\
\midrule
2D-Unet   &  0.916 & 0.106   & 0.853  & 0.118 & 16.76  &  10.79\\
3D-Unet   &  0.925 & 0.041   & 0.865  & 0.065  & 18.51  & 9.86\\
E-CNN   &  0.940  &  0.048   & 0.890  & 0.059  &  12.64  & 7.85\\
OF-Net   &  0.934  &  0.042   & 0.879  & 0.056  &  14.38  & 9.54\\
3D M-Net   &  0.933  &  0.045   & 0.878  & 0.075  &  18.16  & 9.40\\
Ours   &  {\bfseries 0.955}  &  0.030   & {\bfseries 0.916}  & 0.0507  &  {\bfseries 11.49}  & 8.61\\
\bottomrule
\end{tabular}
\label{table:1}
\end{table}

\subsection{Component Analysis}
Table \ref{table:2} presents the segmentation results of our method using individual compartments, where we compared our method without any deformation (3D M-Net), SS network with a single-directional learning (SS-SL), in comparison to the SS-BL.

\begin{table}[h]
\centering
\caption{Segmentation results of our proposed method at individual stages for the 4DCCT dataset.}
\begin{tabular}{*7c}
\toprule
{} &  \multicolumn{2}{c}{Dsc.} & \multicolumn{2}{c}{Jac.} & \multicolumn{2}{c}{HD.}\\
\midrule
{}   & Mean   & Std    & Mean   & Std   & Mean   & Std\\
\midrule
3D M-Net   &  0.933 & 0.045   & 0.878  & 0.075 & 19.16  &  9.40\\
SS-SL   &  0.947 & 0.036   & 0.905  & 0.059  & 12.01   & 8.86\\
Our   &  {\bfseries 0.955}  &  0.030   & {\bfseries 0.916}  & 0.0507  &  {\bfseries 11.49}  & 8.61\\
\bottomrule
\end{tabular}
\label{table:2}
\end{table}

\subsection{Analysis of using SS network with varying time-point intervals}

\begin{figure}[!htbp]
    \centering
    \includegraphics[width=0.5 \textwidth]{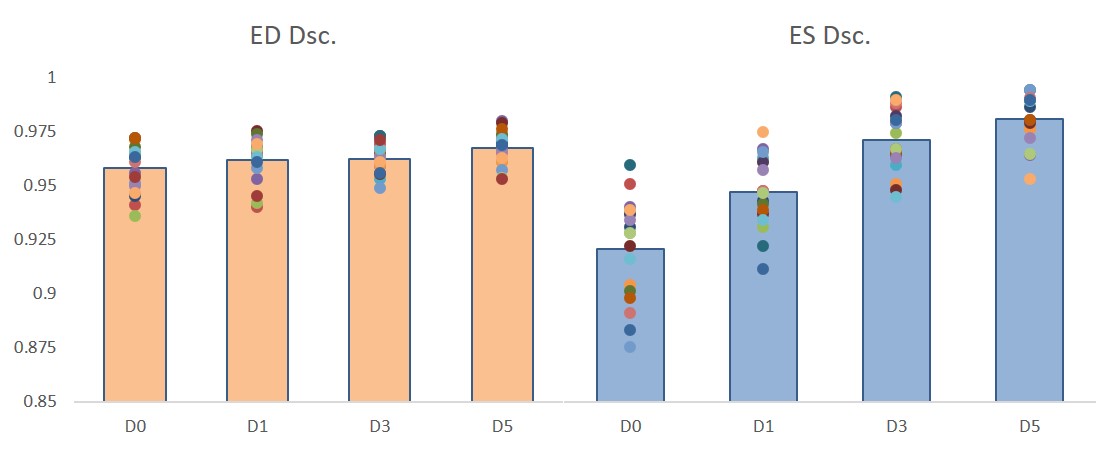}
    \caption{Segmentation results with various deformations under different time intervals.}
    \label{fig:6}
\end{figure}

We evaluated the deformations derived from our SS network with different time-points intervals: (i) D0 – single time-point: no deformation; (ii) D1 – 2 time-points: deformation derived from t1 (ED) and t5 (ES); (iii) D3 – 3 time-points: deformation derived from t1 to t3, and then from t3 to t5 and, (iv) D5 – 5 time-points: deformation derived from 5 adjacent two time points of t1 to t5. Segmentation results were only measured for ED and ES phases since D0 and D1 intervals only used these two phases. 

\section{Discussion}

\begin{figure*}[!htbp]
    \centering
    \includegraphics[width=0.8 \textwidth]{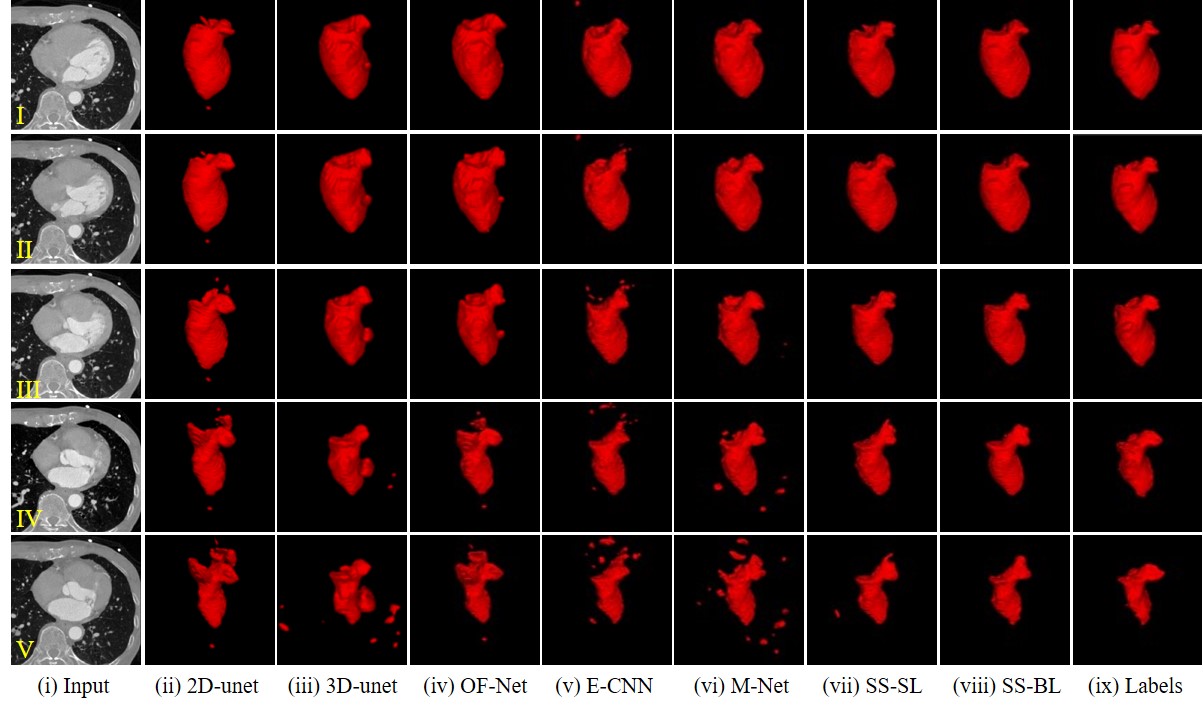}
    \caption{Our 4DCCT dataset segmentation results from a sample patient study showing 5 time-points (every second time-points selected from a total 10 phases) in a cardiac cycle, respectively of 1st to 5th rows. The columns represent the input image, followed by various comparison segmentation results, with SS-SL, SS-BL and the ground truth labels in the last three columns, respectively.}
    \label{fig:7}
\end{figure*}

\begin{table*}[!htbp]
\centering
\caption{Segmentation results of our proposed method at individual stages for the ACDC dataset.}
\begin{tabular}{*7c}
\toprule
{} &  \multicolumn{2}{c}{RVC} & \multicolumn{2}{c}{LVM} & \multicolumn{2}{c}{LVC}\\
\midrule
{}   & Dice   & HD    & Dice   & HD   & Dice   & HD\\
\midrule
3D M-Net & 0.813(0.14) & 27.43(9.61) & 0.804(0.09) & 19.43(9.05) & 0.900(0.08) & 15.94 (7.15)\\
SS-SL & 0.847(0.09) & 16.33 (6.89) & 0.854(0.05) & 12.70 (6.57) & 0.917(0.05) & 10.25 (5.94)\\
\midrule
E-CNN   &  0.907(0.07)  &  13.32(4.85)   & 0.857(0.04)  & 11.47 (8.18)  &  0.930 (0.05)  & 7.96 (6.98)\\
\bottomrule
\end{tabular}
\label{table:3}
\end{table*}

We have identified the following observations with our results: 1) our SS network is able to derive deformation fields between different time-points and this contributed in improving the segmentation performance across the cardiac cycle. This was most evident in challenging ES image phase; 2) we improved performance from our SS-SL network in its inference of the LVC structure from two feature latent spaces (motion and intensity); 3) we further improved the performance with SS-BL network which integrated chronological and reverse-chronological information to alleviate the error predictions and refine the segmentation results and, 4) our SS-BL can be generalized to work with another dataset of 4D MR cardiac image for the segmentation of LVC, RVC and LVM. 

\subsection{Comparison to the Existing Methods}

Our experiments show that our SS-BL resulted in the best overall performance in LVC segmentation when compared to the existing methods. In Table \ref{table:1}, the differences between the 3D U-Net and 2D U-Net demonstrated the benefit of using 3D volume; with 3D convolutional layers enabling improved exploration of the spatial contextual information of the LVC and resulting in better segmentation results. OF-Net used optical flow to model the cardiac motion deformations in 2D space using two time-points (ES and ED phases) of cine MR imaging. In our application of OF-Net to 10 time-point sequence of 4D-C-CT, we noticed that an error from a particular time-point propagated to subsequent time-points, and therefore accumulating in overall errors, as exemplified in Fig. \ref{fig:7}. In contrary, our SS-BL, with its bi-directional motion information derived from only its adjacent time-points, prevented segmentation errors from propagating. We further note that in our SS-BL, by using additional time-points, i.e., from 2 time-points as in OF-Net to 3, and then 5 time-points, we were able to continually improve our segmentation performance as shown in Fig. \ref{fig:6}.

E-CNN \cite{isensee2017automatic} achieved the second best results. Compared with 3D U-Net, E-CNN ensembled the outcomes of 3D U-Net with 2D U-Net, and this, as expected, resulted in improved performance. However, the results were less consistent across the cardiac cycle relative to our proposed method as evident in Fig. \ref{fig:5}. This is expected as with all single time-point methods (U-nets, Isensee), it does not exploit the sequential information that can be used to derive consistency of changes in LVC. Our SS-BL, with the use of bi-directional volumetric deformation, was able to mitigate the variations between the individual time-points. Fig. \ref{fig:7} shows an example where Isensee et al. method which failed to segment the cardiac structures at the end of the systole and resulted in many false positive regions. 

\subsection{Temporal Analysis}

Fig. \ref{fig:6} presents segmentation accuracy with deformation derived from different cardiac cycle time intervals. As expected, segmentation accuracy improved proportionally to the increase in the time-points. We attribute this to the ability to improve the accuracy of deformation field estimation between the phase sequences which are smaller and more relative. This is consistent to the difference in average volumes of the LVC between the phases, where we found volume differences to be greater than 50\% from ED to ES phases (t1 to t5), while there was a maximum of 12\% (based on the LVC volume changes) volume differences among the differences between t1 to t2, t2-t3, t3-t4 and, t4-t5. Nevertheless, even with such large deformation as in t1 to t5 (D1), our method was still able to extract useful deformations to aid in the segmentation (compared to D0). When comparing the ED and ES results, lower performance on ED phase was likely attributed to the fact that ED stage is relatively easier to segment when compared at the ES stage; this is further evident from Fig. \ref{fig:7} among all the segmentation results.

\subsection{Generalizability}
\begin{figure}[!htbp]
    \centering
    \includegraphics[width=0.45 \textwidth]{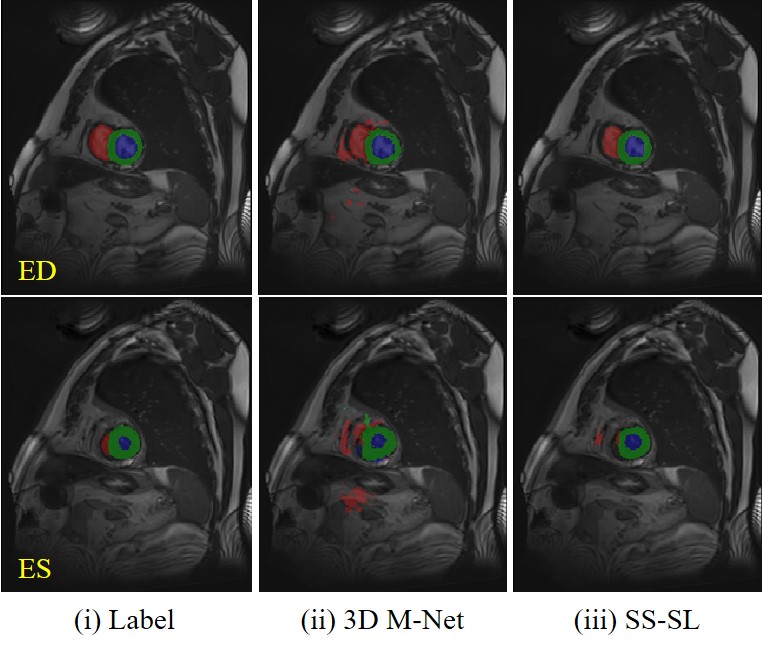}
    \caption{ACDC dataset segmentation results from one example studies in two time-phase in cardiac cycle (end-diastolic and end-systolic).}
    \label{fig:8}
\end{figure}

Although our SS-BL was evaluated specifically for LVC on 4D-C-CT, there is no technical limitations for its generalizability to other cardiac structures to include LVM and RVC, and also to other cardiac imaging modalities such as 4D MR images. To demonstrate this generalizability, we conducted an experiment where we used public ACDC dataset. ACDC is an MR images of the cardiac with only two time-points; it also had lower imaging resolution - the voxel in-plane spacing range from 1.37 to 1.68 mm and inter-slice spacing range from 5 to 10mm. We followed the same approach for pre-processing steps as \cite{jang2017automatic}. Patch size was set to \textit{$224 \times 224 \times 10$} (limited number of slices on z-axis, and thereby allowing for increased size on x- and y-axis when compared to 4D-C-CT). 5-fold cross-validation was applied, consistent to the work in \cite{isensee2017automatic} using the ACDC dataset. Our results are shown in Table \ref{table:3} and Fig. \ref{fig:8}, which has been compared to our baseline model 3D M-Net and the state-of-the-art E-CNN method that achieved the best performance on ACDC.

Our proposed method outperformed 3D M-Net (see Table \ref{table:3} and Fig. \ref{fig:8}), where it consistently improved in the segmentation of RVC, LVM and LVC. However, as expected, it had lower performance compared to the MR image optimized E-CNN. This is because the low correlation of intra-slice context information caused by the large slice thickness of ACDC data, and due to this, it made 2D methods work better in certain images when compared to 3D counterpart for the ACDC data. E-CNN combined the strengths of 2D and 3D methods together to improve the segmentation accuracy. In addition, compared with CT data, the boundary of the cavity in the MR images has less details and the cardiac proportion is smaller compared to CT, which restrict the motion information needed in the SS-BL. Another limitation of applying our SS-BL to ACDC is with it only having two time-points labeled data. SS-BL was designed to exploit the deformation fields across the time-points, and this is a requirement for the SS network to work effectively. We note that E-CNN was designed not to be reliant on the information from the temporal sequence. Despite these drawbacks, we suggest that deformation field from the SS network aids to guide the segmentation on MR cardiac imaging where it outperformed the baseline 3D M-Net \cite{jang2017automatic}. We suggest that our SS-BL can be optimized for MR images by integrating the cardiac motion flow in 2D and 3D which would suitable for low spatial resolution volume images. 

We note the low average segmentation results of ACDC relative to the 4D-C-CT results among all the methods. This is likely due to the lower spatial resolution of the ACDC dataset which restricted the quality of the deformation field derivation that is necessary for our SS-BL. Further, ACDC only had two time-points which restricted our bi-directional learning.

\section{Conclusion}
We proposed a new LVC segmentation method for 4D-C-CT image sequences. By combining deep spatial sequential (SS) network with bi-directional learning (BL) to iteratively learn the temporal and spatial information, our SS-BL method was able to outperform existing methods for cardiac segmentation. In our future studies, we will optimize our algorithm for MR images which usually have lower spatial resolution and distinct cavity boundary compared with CT data. Also, we suggest that our approach is able to expand to other cardiac structure segmentation tasks.

\section*{Acknowledgment}

This work was supported in part by the University of Sydney – Shanghai Jiao Tong University Joint Research Alliance grant and the Australian Research Council (ARC) grant. 
\input{ref.bbl}

\bibliographystyle{IEEEtran}
\bibliography{ref}


\end{document}